\documentclass[%
 aip,
 jmp,%
 amsmath,amssymb,
 reprint,%
]{revtex4-1}

\usepackage{graphicx}% Include figure files
\usepackage{dcolumn}% Align table columns on decimal point
\usepackage{bm}% bold math
\newcommand{\tx}{\textrm}

\input colordvi

\begin{document}

%\preprint{AIP/123-QED}

\title[Applied Physics Letters]{Effects of Interface Roughness Scattering on\\Radio Frequency Performance of Silicon Nanowire Transistors}% Force line breaks with \\
%\thanks{Footnote to title of article.}

\author{SungGeun Kim}
 \email{kim568@purdue.edu.}
\affiliation{
Network for Computational Nanotechnology, Purdue University, West Lafayette, IN 47907, USA
}%

\author{Mathieu Luisier}
\affiliation{
Network for Computational Nanotechnology, Purdue University, West Lafayette, IN 47907, USA
}%
\affiliation{
Integrated Systems Laboratory, Gloriastrasse 35, ETH Z\"{u}rich, 8092 Z\"{u}rich, Switzerland
}%

\author{Timothy B. Boykin}
% \homepage{http://www.Second.institution.edu/~Charlie.Author.}
\affiliation{%
The University of Alabama in Huntsville, Electrical and Computer Engineering, Huntsville, AL 35899, USA
}%

\author{Gerhard Klimeck}%
\affiliation{
Network for Computational Nanotechnology, Purdue University, West Lafayette, IN 47907, USA
}%

\date{\today}% It is always \today, today,

\begin{abstract}
 The effects of an atomistic interface roughness in n-type silicon nanowire transistors (SiNWT) on the radio frequency performance are analyzed. Interface roughness scattering (IRS) is statistically investigated through a three dimensional full--band quantum transport simulation based on the sp$^3$d$^5$s$^*$ tight--binding model. As the diameter of the SiNWT is scaled down below 3 nm, IRS causes a significant reduction of the cut-off frequency. The fluctuations of the conduction band edge due to the rough surface lead to a reflection of electrons through mode-mismatch. This effect reduces the velocity of electrons and hence the transconductance considerably causing a cut-off frequency reduction. %However, the capacitance of the SiNWT is relatively immune to the interface roughness scattering because the total density of states is not changed. It is found that interface roughness scattering reduces the electron velocity more at the beginning of the channel compared to the end of the channel.
%
%Valid PACS numbers may be entered using the \verb+\pacs{#1}+ command.
\end{abstract}

%Use showkeys class option if keyword
\maketitle

 Since the lengths of silicon (Si) metal-oxide-semiconductor field effect transistor (MOSFET) have been scaled down to the sub-100 nm regime, the cut-off frequency has increased significantly to reach hundreds of gigahertz (GHz)\cite{Liou2003,Bennett2005,Schwierz2007}. Even though the cut-off frequency is not the only important parameter in radio frequency (RF) MOSFETs, a high cut-off frequency certainly represents a good criterion for Si MOSFETs to catch up with III-V transistors if other shortcomings are overcome. Power losses due to a long skin depth of the Si substrate, a poor noise figure and a high gate resistance\cite{Burghartz1997} are the examples of such obstacles. Recently there have been tremendous efforts to improve the RF performance of the Si MOSFET and it is becoming competitive to III-V high electron mobility transistor (HEMT)/heterojunction bipolar transistor (HBT) or silicon germanium (SiGe) HBT\cite{Bennett2005,Schwierz2007,VonHaartman2006}.

Silicon-on-insulator (SOI) multi-gate (MG) structures also have been found to be capable of achieving the cut-off frequency predicted by the international technology roadmap for semiconductors (ITRS)\cite{Raskin2006} for RF applications while reducing substrate losses and noise figures\cite{Adan2002}. Gate-all-around (GAA) siicon nanowire transistors (SiNWTs) have attracted attention since it was found that their cut-off frequency can be much larger than that of planar Si MOSFET\cite{Wang2007}.

\begin{figure}[!b]
\includegraphics[width=3.1in]{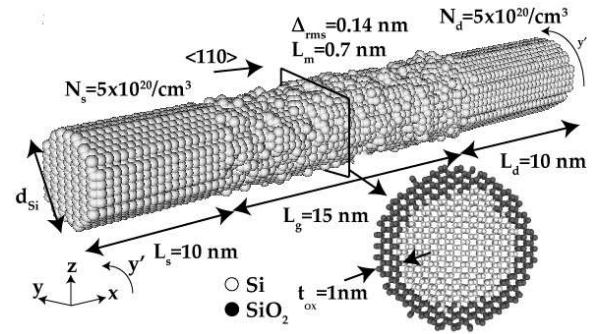}% Here is how to import EPS art
\caption{\label{fig:NWstructure} The simulated silicon nanowire with rough surface in the channel: the root-mean-square roughness height $\Delta_{rms}$ and the correlation length $L_{m}$ are adopted from Ref.~\onlinecite{Goodnick1985}. The crystal orientation $\left<110\right>$ is selected for the electron transport direction. The source/drain doping density $N_s$/$N_d$ is set to $5\times 10^{20} \tx{cm}^{-3}$. The diameter of the nanowire $d_{\tx{Si}}$ varies from 2, 2.5, 3 to 4 nm. The length of the source/drain extension region $L_s$/$L_d$, the gate length $L_g$, and the oxide thickness $t_{\tx{ox}}$ are shown in the figure. The channel of the nanowire is undoped.}
\end{figure}

Traditionally, interface roughness scattering (IRS) has been considered as one of the most important scattering mechanisms. At a high effective electric field, IRS dominates the universal mobility trend\cite{Takagi1994}. In SiNWTs, IRS is still an important scattering mechanism reducing the on-current and the mobility significantly from the ballistic values\cite{SKim2011}.

 This paper focuses on the effects of interface roughness scattering on the RF performance of SiNWTs, especially on the cut-off frequency ($f_T$). For that purpose, a three dimensional full-band quantum transport simulator based on the sp$^3$d$^5$s$^*$ tight-binding (TB) model\cite{Luisier2006,Boykin2004} is used. As the maximum oscillation frequency ($f_{max}$) -- another important figure of merit of the RF MOSFETs is directly related to the cut-off frequency\cite{Wang2007}, the effects of IRS on the theoretical limit of the SiNWT's RF performance can be estimated through this study.

\begin{figure}
\includegraphics[width=3.2in]{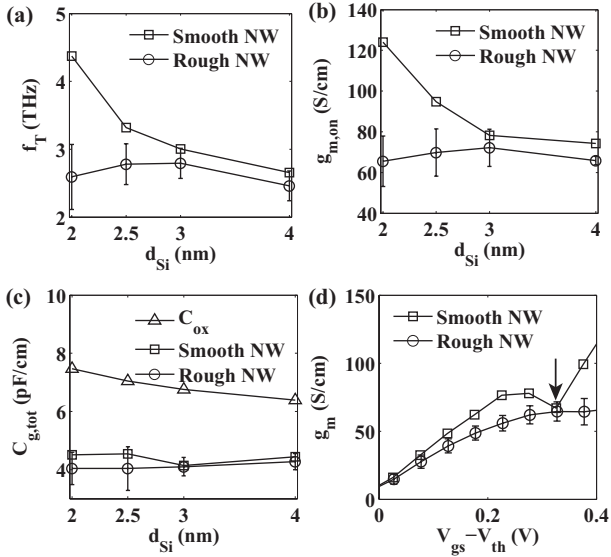}
\caption{(a) The cut-off frequency ($f_{\tx{T}}$), (b) the transconductance ($g_{\tx{m,on}}$), (c) the total gate capacitance $C_{\tx{g,tot}}$ vs diameter $d_{\tx{Si}}$ at the on-state with the gate bias $V_{gs}\sim V_{th}+0.4 V$ with the oxide capacitance $C_{\tx{ox}}=2\pi \epsilon_{\tx{ox}}/\ln\left[2(t_{ox}+d_{\tx{Si}}/2)/d_{\tx{Si}}\right]$ where $\epsilon_{\tx{ox}}$ is the dielectric constant of the oxide, and (d) the transconductance vs gate overdrive for 100 rough nanowire samples (errorbar: standard deviation). All the values except $f_T$ are normalized with the perimeter of the NW.\label{fig:ftvsdia}}
\end{figure}

The structure of the SiNWT studied in this paper is depicted in Fig.~\ref{fig:NWstructure} where the oxide layer is described in the cross-sectional view. The model of the interface roughness in the SiNWT used in the simulation is described in Ref.~\onlinecite{SKim2011} where the influence of the interface roughness scattering on the direct-current (DC) characteristics of SiNWTs is presented. The silicon dioxide (SiO$_2$) layer is included in the transport calculation\cite{SKim2011} to accurately model the wavefunction penetration into the oxide layer.

The cut-off frequency $f_T$ is related to the transconductance $g_{m,on}$ and the total gate capacitance $C_{g,tot}$ through the relationship
\begin{align}
f_T=\frac{g_{m,on}}{2 \pi C_{g,tot}}\label{eq:ft}
\end{align}
where $g_{m,on}$ and $C_{g,tot}$ are calculated through the following expressions at the on-state defined by the gate voltage $V_{gs}$ at $V_{th} + 2/3 V_{dd}$\cite{Chau2005}:
\begin{align}
g_{m,on}=\left.\frac{\partial I_{ds}}{\partial V_{gs}} \right|_{V_{gs}=V_{th}+2/3V_{dd},V_{ds}=V_{dd}}\\
C_{g,tot}=\left. q\frac{\partial N_{1D}}{\partial V_{gs}} \right|_{V_{gs}=V_{th}+2/3V_{dd},V_{ds}=V_{dd}}
\end{align}
where $V_{th}$ is the threshold voltage, $V_{dd}$ the supply voltage and $N_{1D}$ the total electron density under the gate divided by the gate length. The threshold voltage $V_{th}$ is determined using a critical current $I_c=d_{Si}\times10^{-7} (A)$.

The simulated cut-off frequency of a smooth nanowire (NW) is shown in Fig.~\ref{fig:ftvsdia}(a). The results obtained here are similar to the data calculated in Ref.~\onlinecite{Wang2007}. The cut-off frequency increases as the nanowire diameter decreases. This is to first order a consequence of the improvement of the injection velocity in a $\left<110\right>$ silicon nanowire (SiNW) with smaller diameter\cite{Phytos2008Elec}.

\begin{figure}[t]
\includegraphics[width=3.6in]{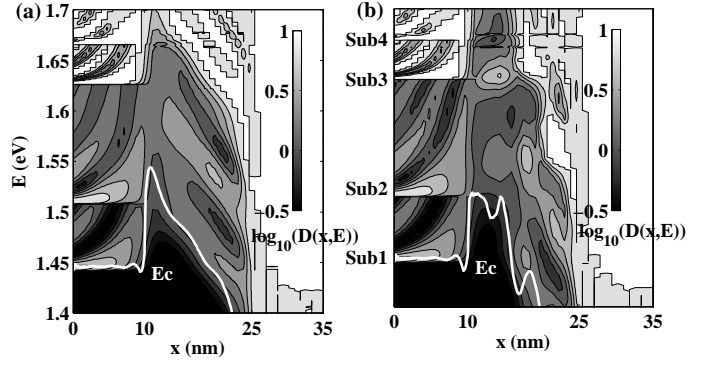}
\caption{The density of states $D(x,E)$ in a logarithmic scale for (a) a smooth NW or (b) a rough NW resolved in the transport axis $x$ and the energy $E$ near the on-state with $V_{gs}\sim V_{th}+0.4 V$. The channel of the NWFET starts from $x=10$ nm and extends to $x=25$ nm.\label{fig:DOS2D}}
\end{figure}

As shown in Fig.~\ref{fig:ftvsdia}(b), the transconductance $g_{m,on}$ is reduced significantly by the IRS while $C_{g,tot}$ is not affected much (Fig.~\ref{fig:ftvsdia}(c)). The reduction of $g_{m,on}$ is due to reflections caused by the rough interface. A small dip in $g_m$ marked by an arrow in Fig.~\ref{fig:ftvsdia}(d) is an indication that the second subband starts to carry the current~\cite{RSKim2008}. In rough NWs, this dip is smoothed out due to subband mixing.%As the gate bias increases, more and more subbands start to carry the currents. %At a low temperature, the transconductance increases in steps as the gate voltage increases. This is a signature of 1D transport\cite{RSKim2008}.

%However, the steps are smoothed out when the temperature is increased to 300 K due to thermal broadening. A close look at the simulation results in Fig.~\ref{fig:ftvsdia}(d) for the smooth nanowire reveals that there is still a trace of this step-like behavior at $V_{\tx{gs}}-V_{\tx{th}} \sim 0.35$ V.

%Fig.~\ref{fig:DOS2D}(a) shows the density of states (DOS) in a smooth NW where a clear distinction between subbands can be seen in the channel. One can observe that the subbands in the channel are not aligned in energy throughout the NW. This is another reason why the step is not clear as in a low drain bias (e.g. see Ref.~\onlinecite{RSKim2008}) in addition to the thermal broadening of the Fermi function in the source. However, in rough nanowires the subbands are mixed in the channel related to the fluctuations of the conduction band edge as shown in Fig.~\ref{fig:DOS2D}(b). The distinction between subbands disappears. As a result, the step-like behavior also disappears in $g_m$ of rough nanowires shown in Fig.~\ref{fig:ftvsdia}(d) and hence the turn-on of the second subband is not clear in rough nanowires.

\begin{figure}
\includegraphics[width=3.4in]{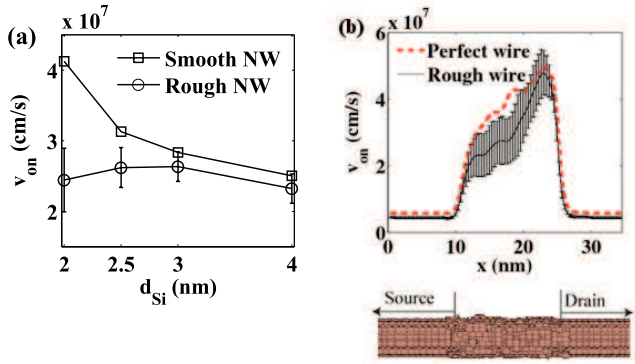}
\caption{(a) The average electron velocity at the on-state for NWs with a different diameter and (b) the electron velocity along the channel at the on-state for rough NWs with diameter 2 nm (errorbar: standard deviation).\label{fig:velocity}}
\end{figure}

Mismatches of the subbands throughout the channel of the rough nanowire also can be observed in Fig.~\ref{fig:DOS2D}(b). This causes reflections of electrons causing reduction of the electron velocity which, in turn, reduces the transconductance. Fig.~\ref{fig:velocity}(b) shows the electron velocity throughout the smooth NW and the rough NWs with the diameter 2nm. The electron velocity is significantly reduced by interface roughness scattering.

 One thing noticeable in Fig.~\ref{fig:velocity}(b) is that the IRS causes a reduction of the electron velocity at the beginning of the channel, but not much at the end of the channel. Electrons gain a relatively large kinetic energy due to a large electric field at the end of the channel. As a result, the fluctuation of the conduction band edge at the end of the channel does not affect the electron velocity significantly.

The cut-off frequency relationship (Eq. \ref{eq:ft}) can also be expressed as
\begin{align}
f_T=\frac{v_{on}}{2 \pi L_{g}}\tx{.}\label{eq:ft_vel}
\end{align}
Therefore, the average electron velocity $v_{on}$ can be calculated from the cut-off frequency. The transit time under the gate $\tau_T$ is determined from $v_{on}/L_g$, such that the average velocity is an effective velocity with which electrons flow in the channel when a small signal is applied to the gate. As shown in Fig.~\ref{fig:velocity}(a), it turns out that $v_{on}$ is higher than the ballistic injection velocity ($\sim 1.5\times 10^7 \tx{cm/s}$ from Ref.~\onlinecite{Liu2008}) because electron velocity is not saturated in the beginning of the channel as in a long channel transistors. It can be observed that the average electron velocity $v_{on}$ is close to the velocity in the middle of the channel.

The total gate capacitance is also an important parameter in the SiNWT. Experimentally it is found that the total gate capacitance is reduced from the oxide capacitance due to volume inversion of carriers in a nanowire\cite{Suk2007}. In the simulated NW, the total gate capacitance is found to be much smaller than the oxide capacitance as shown in Fig.~\ref{fig:ftvsdia}(c). %Moreover, the rough surface further squeezes the SiNW because the smallest cross-section in the channel of a rough NW is a bottle-neck of the carrier transport in the nanowire.

\begin{figure}[t]
\includegraphics[width=3.4in]{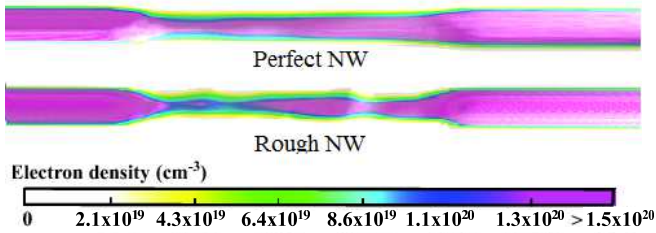}
\caption{The electron density from source to drain for (top) the smooth NW and (bottom) the rough NW (the same sample selected for Fig.~\ref{fig:DOS2D}) at the on-state with $V_{gs}\sim V_{th}+0.4 V$.\label{fig:eden}}
\end{figure}

Fig.~\ref{fig:eden} shows the electron density along the NW where it can be observed that the electron density is fluctuating throughout the channel as compared to the smooth NW. Interface roughness causes mode mixing and additional reflections in the current.  It does, however, not modify the total density of states (DOS), and therefore the capacitance of the nanowire. Therefore, the total gate capacitance is relatively unaffected by rough interfaces.%The overall reduction of the effective cross-section for the electron transport due to the rough interface increases the effective oxide thickness and reduces the total gate capacitance.
%However, this effect is much smaller than the reduction of the transconductance and hence the overall cut-off frequency is reduced as a result.

In conclusion, the cut-off frequency of SiNWTs is statistically studied through quantum transport simulation using a realistic modeling of the rough Si/SiO$_2$ interface. It is found that the rough surface causes back-scattering and reduces the velocity of electrons via modifying the DOS in the channel. Mode-mismatch due to interface roughness scattering reduces the overall transconductance, but does not significantly affect the total gate capacitance. In addition to the cut-off frequency degradation, its variability is another issue that should be addressed in RF SiNWTs.

We acknowledge Materials, Structures and Devices Focus Center, one of the six research centers funded under the Focus Center Research Program (a Semiconductor Research Corporation entity); Rosen Center for Advanced Computing, National Center for Computational Sciences, National Institute for Computational Sciences, and Texas Advanced Computing Center for the supercomputing resources; and NanoHUB for the computational resources.

\end{document}